\documentstyle[aps,epsf]{revtex}
\def\beq{\begin{equation}}
\def\eeq{\end{equation}}
\def\bea{\begin{eqnarray}}
\def\eea{\end{eqnarray}}
\def\nn{\nonumber}
\def\ba{\begin{array}}
\def\ea{\end{array}}   

\begin{document}
\title{Phenomenology of Neutrino Oscillations
\footnote{Plenary talk given at WHEPP-6, 3-16 January 2000,  Chennai, India.}}
\author{G. Rajasekaran}
\address{ Institute of Mathematical Sciences, Chennai 600113 }
\maketitle
\begin{abstract}
The phenomenology of solar, atmospheric, supernova and
laboratory neutrino oscillations is described. Analytical formulae for
matter effects are reviewed. The results from oscillations are confronted with
neutrinoless double beta decay.   
\end{abstract}

\section{Introduction}
In this talk, I shall try to give a bird's eye view of the current
status of neutrino oscillations.  Also I shall highlight local work
wherever possible since phenomenological contributions have been made
from the Institute of Mathematical Sciences on various aspects of
neutrino oscillations.

Solar neutrino physics is 30-years old. Starting from the pioneering
$Cl$ experiments of Davis and collaborators \cite{davis1}, all the
experiments have observed the depletion of the solar neutrinos as
compared to the theoretically calculated flux from the standard solar
model
 \cite{bahcall1}. Atmospheric neutrino physics is much younger. The
observation of the reduction in the
 $\nu_\mu/\nu_e$ ratio is only 10-years old. Nevertheless, atmospheric
neutrino physics has won over solar neutrino physics, since the
evidence for neutrino oscillation from the former is derived from
ratios and is hence independent of the absolute flux. On the whole,
very good evidence for oscillation exists from both fronts. There are
still loose ends ; for instance, the recoil energy spectrum of the
solar neutrinos does not agree with the theoretically calculated one,
even with oscillations.

Alternative ideas \cite{pakvasa} must still be explored and ruled out,
before neutrino oscillations are finally established as a part of
Physics. Nevertheless it is fair to state that oscillations are the
natural explanation for the observed neutrino anomalies.  For,
oscillation is the most conservative explanation for the anomalies. It
does not violate any known principle of physics and it does not invoke
any exotic new physics. It only uses ordinary principles of quantum
mechanics, especially the principle of superposition. In quantum
mechanics, the neutrino states will mix and oscillate, if they have
masses.

Hence, I think it is time to look at the next job. Attention must now
shift from the oscillation itself, to determining the 6 fundamental
parameters (2 mass differences, 3 mixing angles and one CP violating
phase) describing the oscillation phenomena among the 3 flavours
$\nu_e, \nu_\mu$ and $\nu_\tau$. Even after 30 years, none of these
parameters is yet determined with any certainty. But, progress has
recently been made in limiting their ranges or determining their orders
of magnitude.  One parameter, namely the mixing angle $\phi$, has been
bracketed rather closely.

It is important to stress the desirability of explaining neutrino
anomalies and fitting the data using a three-neutrino framework. For,
there exist three neutrinos. The high-quality data coming out from the
great neutrino experiments of the present-day deserve to be treated by
a realistic $3-\nu$ analysis rather than the $2-\nu$ toy model often
used by the experimenters to give their results.  This is the point of
view with which we started our $\nu$-phenomenology work at IMSc
\cite{mohan1}. When we started, there were very few groups
doing a $3-\nu$ analysis.

Having said the above, I have also to point out that Nature has helped
the (lazy) $2-\nu$ people.  First, it chose the mass hierarchy :
\beq
m^2_2 - m^2_1 \quad << \quad  m^2_3 - m^2_2    \label{hierchy}
\eeq
so that solar $\nu$ could be explained by $\delta m^2_{21} ( \equiv
m^2_2 - m^2_1)$ and atmospheric $\nu$ could be explained by $\delta
m^2_{32} (\equiv m^2_3 - m^2_2)$.  Next, the CHOOZ reactor experiment
showed that one of the mixing angles $(\phi)$ is so small that the
$3-\nu$ problem decouples into two effective $2-\nu$ problems.

Although neutrino oscillation is a direct consequence of quantum
mechanics, it leads to a result of profound consequence for Physics and
Astrophysics -- the result that neutrinos have mass. That is the
importance of the whole subject of neutrino oscillations.
Neutrino mass is the only concrete evidence we have for physics beyond
the Standard Model (SM) of high energy physics. Every other physics
beyond the SM, that is being searched for, has remained speculative so
far. Every other experimental signal for physics beyond the SM that
appears now and then on the horizon has been disappearing in about 6
months to one year. Neutrino mass is the only evidence for physics
beyond SM that has remained robust for the past 30 years.

\section{Results}
The neutrino flavour states $|\nu_\alpha \rangle (\alpha = e, \mu,
\tau)$ are linear superpositions of the neutrino mass eigenstates
$|\nu_i\rangle  (i=1,2,3)$ with masses $m_i$ :
$
|\nu_\alpha \rangle   =  \Sigma_i U_{\alpha i} |\nu_i\rangle 
$
where $U$ is the $3 \times 3$ unitary matrix :
\beq
U  =  \pmatrix{c_\phi c_\omega & c_\phi s_\omega & s_\phi \cr
-c_\psi s_\omega -s_\psi s_\phi c_\omega e^{i\delta} & 
c_\psi c_\omega   -s_\psi s_\phi s_\omega e^{i\delta} & 
s_\psi c_\phi   e^{i\delta} \cr
s_\psi s_\omega - c_\psi s_\phi c_\omega e^{i\delta} &
-s_\psi c_\omega -c_\psi s_\phi s_\omega e^{i\delta} &
c_\psi c_\phi e^{i\delta}}        \label{uvac}
\eeq
where $c$ and $s$ stand for sine and cosine of the angle
appearing as subscript.

From the oscillation phenomena, one has to determine the 6 parameters
\linebreak $\delta m^2_{21}, \delta m^2_{32}, \omega, \psi, \phi$ and $\delta$.
Most of the presently studied oscillation phenomena are insensitive to
the CP-violation parameter $\delta$. So let us ignore $\delta$ but we
shall justify it at the end of the section.

Under the hiearchy assumption of Eq.(\ref{hierchy}), one can show that
the solar neutrino problem depends only on $\delta m^2_{21}, \omega $
and $\phi$ while the atmospheric neutrino problem depends only on
$\delta m^2_{32}, \psi $ and $\phi$. It is this simplification that
allows us to analyse the two problems within a $3-\nu$ framework under
reasonable control and one gets a fairly broad and stable set of
allowed regions in the 5-parameter space
\cite{mohan1,mohan2}.
If we temporarily put $\phi$ as zero, then the two
problems decouple and the results are the following :-

\noindent There are three solutions of the solar neutrino problem :

\vskip .5cm

1.   MSW-small angle  :  $\delta m^2_{21} \approx 10^{-5} eV^2,
sin^2 2\omega \approx 10^{-3}$ \\

2.   MSW-large angle :   $\delta m^2_{21} \approx 10^{-5} eV^2,
sin^2 2\omega \approx 1$ \\

3. Vacuum oscillations :  $\delta m^2_{21} \approx 10^{-10} eV^2,
sin^2 2\omega \approx 1$ 

\noindent Atmospheric neutrinos problem has the solution :
$
\delta m^2_{32} \approx 10^{-3} eV^2, \ sin^2 2\psi \approx 1
$

The nonobservation of any depletion of $\bar{\nu}_e$ in the 
reactor experiment at CHOOZ \cite{chooz} turns out to be a crucial
result.  Interpreted in the $3-\nu$ framework \cite{mohan3}, with the
hieararchy assumption of Eq.\ref{hierchy} it implies $ \phi \ < 9^o$,
which is a very powerful constraint and this result justifies the
neglect of $\phi$ in the analysis of solar and atmospheric $\nu$. This
constraint is independent of CP violation.

One can show that CP violation always occurs
in the combination $\sin\phi e^{\pm i\delta}$. So, in view of the CHOOZ
result, CP violation is suppressed in all neutrino oscillation phenomena.

In the next few sections, we shall briefly describe how these results
were obtained.

\section{Oscillations in vacuum and matter}
The probability of a neutrino of flavour $\alpha$ to be observed with
flavour $\beta$ after a distance of travel $L$ in vacuum is given by
\beq
P(\nu_\alpha \rightarrow \nu_\beta) \ = \ \delta_{\alpha \beta } - 4
\sum_i \sum_{j > i} U_{\alpha i} U_{\beta i} U_{\alpha j} U_{\beta
j}  \sin^2 \left[ \frac{1.27 \delta m^2_{ij} L}{E}\right] \label{vactrans}
\eeq
where $U$ is defined in Eq.(\ref{uvac}), $\delta m^2_{ij}$ is in $eV^2$, $L$ is
in metres, neutrino energy $E$ is in $MeV$ and $CP$ violation is ignored. 

In matter (especially of varying density), the above formulae are
drastically changed because of the famous Mikheyev-Smirnov-Wolfenstein
(MSW) effect. We consider the propagation of the neutrinos through solar
matter. Let a neutrino of flavour $\alpha$ be produced at time $t = t_o$
in the solar core. Its state vector is
$
|\Psi_\alpha (t_o)\rangle   =  |\nu_\alpha \rangle   =  \sum_i
U^c_{\alpha i} | \nu^c_i \rangle    
$
where $|\nu^c_i\rangle $ are the mass eigenstates with mass eigenvalues
$m^c_i$ and mixing matrix elements $U^c_{\alpha i}$ in the core of the
sun. The neutrino propagates in the sun adiabatically upto $t_R$ (the
resonance point),  makes nonadiabatic Landau-Zener transition $i
\rightarrow j$ at $t_R$ with probability amplitude $M^{LZ}_{ji}$,
propagates adiabatically upto $t_1$ (the edge of the sun) and propagates
as a free particle upto $t_2$ when it reaches the earth.

The state vector at $t_2$ is
\beq
|\Psi_\alpha(t_2) \rangle  \ = \ \sum_{i,j} |\nu_j \rangle  M^{LZ}_{ji}
U^c_{\alpha i} exp \left( -i
\int^{t_2}_{t_R} \epsilon_j (t) dt -i \int^{t_R}_{t_o} \epsilon_i(t)
dt \right)
						\label{endstate}
\eeq
where $\epsilon_i(t) \left\{ = E + m^2_i(t)/2E\right\}$ are
the matter-dependent energy eigenvalues in the sun upto
$t_1$ and vaccuum eigenvalues for $t_1 < t < t_2$. The
probability for detecting a neutrino of flavour $\beta$ on
the earth is
\bea
| \langle  \nu_\beta | \Psi_\alpha (t_2) \rangle  |^2 & =& 
\sum_{iji'j'} U^*_{\beta j} U_{\beta j'} M^{LZ}_{ji}
M^{LZ^*}_{j'i'} U^c_{\alpha i} U^{c^*}_{\alpha i'}  \nn \\  
  & & exp \left\{ -i \int^{t_2}_{t_R} (\epsilon_j - \epsilon_{j'}) dt 
      -i \int^{t_{R}}_{t_o} (\epsilon_i - \epsilon_{i'} dt\right\}
					\label{detprob}
\eea
Next comes the crucial step \cite{parkes,kuo1} of averaging over $t_0$
and $t_2$ and the assumption that the oscillations are rapid enough so
that the averaged exponential in this equation can be replaced by
$\delta_{ii'} \delta_{jj'}$. Calling this averaged probability as
$P^D_{\alpha \beta}$ (The probability for a $\nu_{\alpha}$ produced in
the sun to be detected as a $\nu_\beta$ on the earth at daytime), we
get
\beq
P^D_{\alpha \beta} = \sum_{ij} |U_{\beta j}|^2 |M^{LZ}_{ji}|^2 
                     |U^c_{\alpha i}|^2       \label{avgprob}
\eeq

\section{Solar, atmospheric and reactor neutrinos}
Extensive literature exists on the solar \cite{mohan1,fogli1} and 
atmospheric neutrinos \cite{mohan2,fogli2}. So, we shall be brief.
 
The experimental results on the total solar neutrino flux from the
three types of detectors ($Cl,~Gl$ and $H_2O$) are the following
\cite{davis2}:  $R_{Cl}=0.33\pm 0.028; ~R_{Ga} = 0.56 \pm 0.05;
~R_{SK}= 0.475 \pm 0.015 $ where we have given the ratios of the
experimental rates to the theoretical rates without neutrino
oscillations calculated in the Standard Solar Model (SSM). One can see
that the statistical uncertainty is the least for the SuperKamioka (SK)
water Cerenkov detector, which is thus presaging the era of precision
neutrino physics. Since the three types of detectors are sensitive to
different regions of the solar neutrino spectrum, the above three
numbers already contain some spectral information. Ascribing the above
observed depletion factors to oscillation and using the formulae in
Eq.(\ref{vactrans}) or (\ref{avgprob}), the best fits for the neutrino
parameters can be obtained, with the results already discussed in
Sec.2.  However, the recoil electron energy spectrum as measured by SK
does not fit \cite{bahcall2} with any oscillation scenerio, at the
higher energy end where the observed number is too large (but with
large errors).

Since the neutral current (NC) week interaction is flavour-blind, the
flux of solar $\nu$ detected through NC node would be the total
$(\nu_{e} + \nu_\mu + \nu_\tau)$ flux and hence is independent of
oscillation and would be a test of the standard solar model. Also, the
ratio CC/NC, would be a test of neutrino oscillation independent of the
uncertainties of the solar models. Therefore great expetations have
been raised by the SNO detector \cite{donald} which will soon give the
first results.  Another exciting avenue will open when Borexino
\cite{raghavan1} starts its operation, since it is the first detector
zeroing in on the monochromatic $\nu$ line from the $Be^7$ decay in the
sun.

Cosmic rays impingent on the atmosphere produce hadrons
(especially pions) that decay, resulting in a wide spectrum
of $\nu_\mu, {\bar \nu}_\mu, \nu_e, {\bar \nu}_e$ ranging in
energy upto about 100 GeV, with the calculated ratio
$R_{cal} = (N_{\nu_\mu} + N_{{\bar \nu}_\mu})/(N_{\nu_e} +
N_{{\bar \nu}_e})$ approximately equal to $2$, whereas the
experimentally measured ratio is nearer to unity. More
precisely, $r$ defined as $R_{obs}/R_{cal}$ is about 0.6.
This is the atmospheric neutrino problem whose solution is
the oscillation of $\nu_\mu (\bar{\nu}_\mu)$ into
$\nu_\tau(\bar{\nu}_\tau)$.

Abundant data from superkamioka detector \cite{fukuda} is now available
on the zenith angle dependence as well as on the
energy-distribution of the $\nu_\mu$ and $\nu_e$ events.
All these are consistent with $\nu_\mu$ oscillating into
$\nu_\tau$ over the distance scale of about 10,000 km ; so
it is for the upward-going $\nu_\mu$ travelling through the
earth that the effect is most dominant. The effect of earth
matter in the atmospheric neutrino problem is not
significant \cite{mohan2} at the present level of accuracy and hence
one can use Eq.(\ref{vactrans}). The resulting neutrino
prameters were given in Sec 2.

Although the analysis is performed in terms of ratios such
as $\nu_\mu/\nu_e$ or up/down and hence is relatively
insensitive to the rather large uncertainties that exist in
the primary cosmic ray flux and spectrum,
further improvement in our knowledge of the latter will be
essential for the complete understanding of all atmospheric
neutrino data.

Also, it is worth pointing out that, if the above
explanation of atmospheric neutrino anomaly is correct,
then the upward-going atmospheric neutrino beam has the approximate
composition of $\nu_e : \nu_\mu : \nu_\tau = 1 : 1:1$.
The direct detection of this $\nu_\tau$ through
production of $\tau$ will be a crucial test. 

Nuclear fission reactors are powerful sources of
$\bar{\nu}_e$. The detector was placed about 1
km away from the CHOOZ power reactor \cite{chooz} and the observed
$\bar{\nu}_e$ flux was compared with the calculated flux :
$
\phi_{obs}/ \phi_{cal}  =  0.98 \pm 0.04 \pm 0.04
$. 
If $\delta m^2_{32} >> \delta m^2_{21}$ and $\delta
m^2_{21} \stackrel{\sim}{<} 10^{-5} eV^2$, the $3-\nu$
formula of Eq.(\ref{vactrans}) reduces \cite{mohan3} to the 2-$\nu$ formula 
\beq
P(\bar{\nu}_e \rightarrow \bar{\nu}_e) \ = \ 1 - \sin^2 2
\phi \sin^2 \frac{1.27 \delta m^2_{32} L}{E}    \label{nubarsurv}
\eeq
Comparison with the CHOOZ result
yields $\phi < 12^o$ for $\delta m^2_{32} \stackrel{>}{\sim} 10^{-3} eV^2$. 
CHOOZ have now 
improved their limit to $\phi < 9^o$.

For the first time, a negative result on
neutrino oscillations from laboratory experiment has given
a constraint of significance in the context of solar and
atmospheric neutrinos. That is the importance of the CHOOZ
experiment. Independent confirmation of this result has
come from the Palo Verde experiment, although with less
statistics. Hopefully the dependence on the calcualted flux
$\phi_{cal}$ will be removed in the future by placing
another detector near the reactor.

\section{Neutrinos through the earth and the moon}
Neutrino oscillation is a complex phenomenon depending on many unknown
parameters (six parameters for three flavours $\nu_e, \nu_\mu$ and
$\nu_\tau$) and a considerable amount of experimental work and
ingenuity will be required before the neutrino problem is solved.

Whenever physicists are confronted with a beam of unknown properties,
they pass it through different amounts of matter. Nature has
fortunately provided us with such opportunities (See Fig.1) : (a)
Neutrinos produced in the solar core pass through solar matter ; (b)
solar neutrinos detected at night pass through earth ; (c) solar
neutrinos detected during a solar eclipse pass through the moon ; (d)
solar neutrinos detected at the far side of earth during a solar
eclipse pass through the moon and earth ; (e) upward going atmospheric
neutrinos pass through the earth. To these we may add two more
experiments of the future : (f) Long-base-line experiments of
accelerator and reactor produced neutrinos and (g) detection of
geophysical neutrinos \cite{geophysical}.

\begin{figure}[htbp]
\epsfxsize=12cm
\centerline{\epsfbox{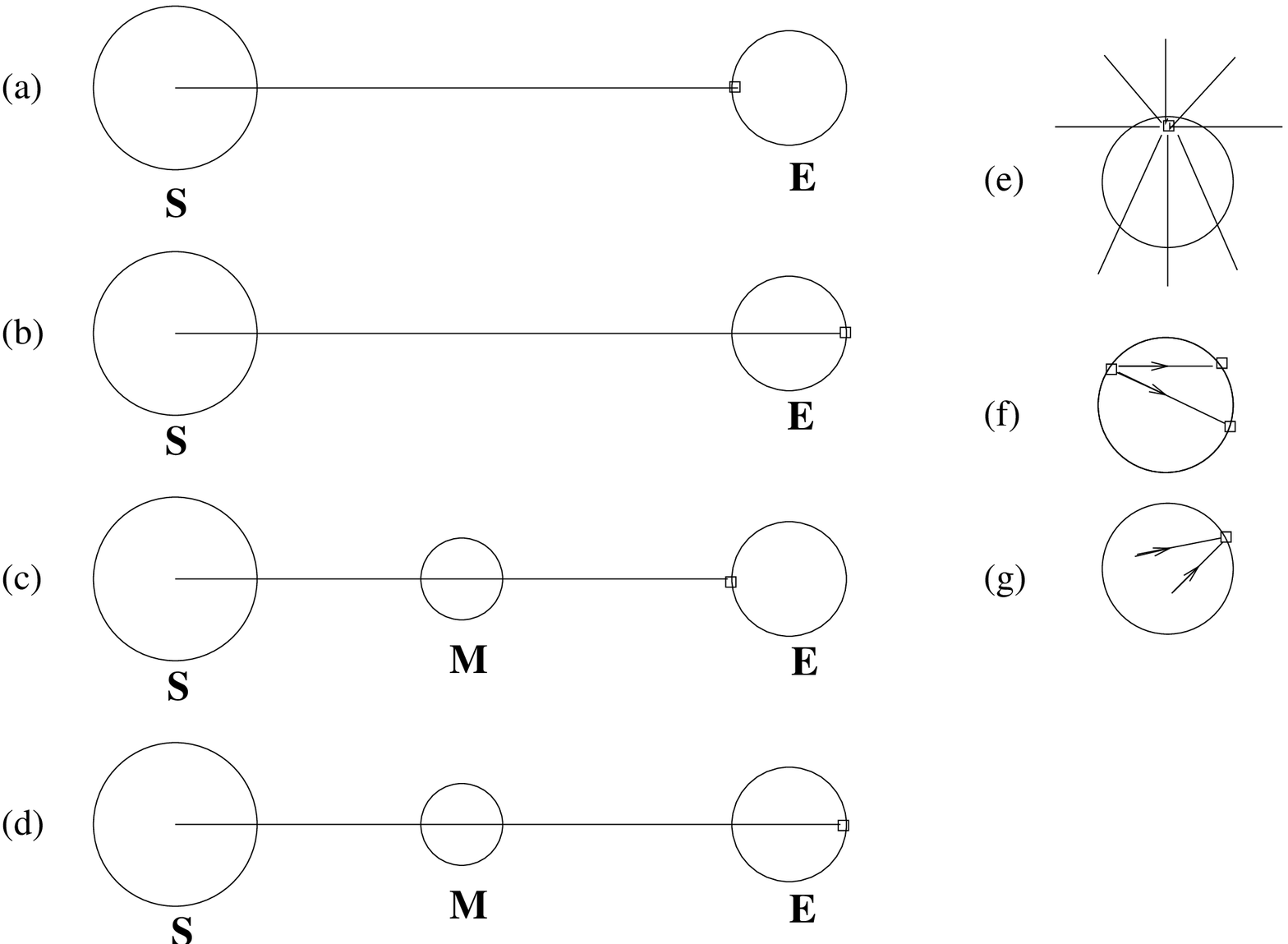}}
\caption{Neutrinos passing through various pieces of matter}
\label{fig:eclipse}
\end{figure}

It is possible to treat these effects analytically. The analytical
formula for (a) was already given in Eq.(\ref{avgprob}). We shall now
derive the formulae for (b) the night effect \cite{mohan4}, (c) the
eclipse effect \cite{mohan5} and (d) the double eclipse effect
\cite{mohan5}.

\subsection{The night effect}

Starting with $|\Psi_\alpha(t_2)\rangle $ on the surface of the earth
given by Eq.(\ref{endstate}), we multiply the righthand side by $\sum_k
|\nu^E_k \rangle  \langle \nu^E_k|(=1)$ where $|\nu^E_k\rangle
(k=1,2,3)$ is the complete set of matter dependent mass eigenstates
just inside the earth. If the neutrino propagates adiabatically upto
$t_3$ on the other side fo the earth (we shall soon correct for
nonadiabatic jumps during the propagation), the state vector at $t_3$
is
\bea
|\Psi_\alpha (t_3)\rangle  &=&  \sum_{i,j,k} |\nu^E_k \rangle \langle  
      \nu^E_k |\nu_j\rangle M^{LZ}_{ji} U^c_{\alpha i} \nn \\
      & &exp \left\{-i \int^{t_3}_{t_2} \epsilon_k dt 
      -i \int^{t_2}_{t_R} \epsilon_j dt 
      -i \int^{t_R}_{t_0} \epsilon_i dt \right\}  \label{nightevo}
\eea
This expression automatically contains $\langle  \nu^E_k |\nu_j\rangle$ 
which is the
probability amplitude for nonadiabatic transition $j \rightarrow k$ 
at the vacuum-earth
boundary and we shall call it $M^E_{kj}$ :
\beq
M^E_{kj} = \langle  \nu^E_k | \nu_j \rangle = \sum_\sigma \langle  \nu^E_k
        |\nu_\sigma \rangle   \langle  \nu_\sigma | \nu_j \rangle 
      = \sum_\sigma U^E_{\sigma k} U^*_{\sigma j}   \label{nonad}
\eeq
where $\nu^E_k$ and $U^E$ are mass eigenstates and mixing matrix just
inside the earth.

Averaging the probability $|\langle  \nu_\beta | \Psi_\alpha
(t_3)\rangle  |^2$ over $t_R$ results in the desired incoherent mixture
of mass eigenstates of neutrinos reaching the surface of the earth.
Calling this average probability as $P^N_{\alpha \beta}$ (the
probability for $\nu_\alpha$ produced in the sun to be detected as
$\nu_\beta$ in the earth at night), we can write the result as

\beq
P^N_{\alpha \beta} = \sum_j P^S_{\alpha j} P^E_{j\beta}  \label{nightprob}
\eeq
where $P^S_{\alpha j}$ is the probability of $\nu_\alpha$ produced in
the sun being detected as $\nu_j$ (mass eigenstate) as it enters the
earth and $P^E_{j\beta}$ is the probability of $\nu_j$ entering the
earth to be detected as $\nu_\beta$ after it propagates through the
earth. These are given by
\beq
P^S_{\alpha j} = \sum_i |M^{LZ}_{ji} |^2 |U^c_{\alpha i}|^2  \label{ps}
\eeq
\beq
P^E_{j\beta} = \sum_{k,k'} U^{E^*}_{\beta k} U^E_{\beta k'} M^E_{kj} M^E_{k'j} 
   exp (-i \int^{t_3}_{t_2} (\epsilon_k - \epsilon_{k'}) dt)  \label{pe}
\eeq
It is important to note that the factorization of probabilities in
Eq(\ref{nightprob}) (which has been derived here as a consequence of
the averaging over $t_R$), is valid only for mass eigenstates in the
intermediate state. An equivalent statement of this result is that the
density matrix is diagonal only in the mass-eigenstate representation
and {\it not} in the flavour representation.

During the day, put $t_3 = t_2$ so that $P^E_{j\beta}$ becomes
$|U_{\beta j}|^2$ and so Eq.(\ref{nightprob}) reduces to
Eq.(\ref{avgprob}). One can justify \cite{parkes} the averaging over
$t_0$ and $t_2$ by the facts that the neutrinos are produced over an
extended region in the solar core and they are detected over an
extended region or time since the detector is moving with the earth.
While averaging over $t_0$ and $t_2$ is equivalent to averaging over
$t_R$ as far as $P^D_{\alpha \beta}$ is concerned, it is not so for
$P^N_{\alpha \beta}$, but we have adopted the latter method for
$P^N_{\alpha \beta}$ because of its simplicity in giving us the
factored probability expression in Eq(\ref{nightprob}).

However, two points have to be made : (i) For $P^N_{\alpha \beta}$, it
is not justified to average over $t_2$ or $t_3$ since we would like to
detect the neutrinos during a narrow time-bin in the night. (ii)
Averaging over $t_R$ (as we have done) may be partially justified since
the result may be effectively the same for energy-integrated rates.
However, this argument does not apply for Borexino \cite{raghavan1}, where
the monochromatic $Be^7$ neutrino line spectrum will be detected.

Next we show how to take into account nonadiabatic jumps
during the propagation inside the earth. Consider
$\nu$ propagation through a series of slabs of matter,
density varying inside each slab smoothly but changing
abruptly at the junction between adjacent slabs. The state
vector of the neutrino at the end of the $n^{th}$ slab
$|n\rangle $ is related to that at the end of $(n-1)$th slab
$|n-1\rangle $ by $|n\rangle  = F^{(n)} M^{(n)} |n-1\rangle$
where $M^{(n)}$ describes the nonadiabatic jump occuring at
the junction between $(n-1)$th and $n$th slabs while
$F^{(n)}$ describes the adiabatic propagation in the $n$th
slab. They are given by 
\beq
M^{(n)}_{ij} = \langle  \nu^{(n)}_i | \nu^{(n-1)}_j\rangle  =
(U^{(n)^\dagger} U^{(n-1)})^*_{ij} \label{Mn} 
\eeq
\beq
F^{(n)}_{ij} = \delta_{ij} exp \left( -i\int^{t_n}_{t_{n-1}} 
              \epsilon_i (t) dt\right)     \label{Fn}
\eeq
where the indices $(n)$ and $(n-1)$ ocuring on $\nu$ and
$U$ refer respectively to the $n$th and $(n-1)$th slabs at
the junction between these slabs. Also note that $M^{(1)}$
is the same as $M^E$ defined in Eq.(\ref{nonad}). Defining the
density matrix at the end of the $n$th slab as $\rho^{(n)}
= |n\rangle \langle n|$, we have the recursion formula
\beq
\rho^{(n)} = F^{(n)} M^{(n)} \rho^{(n-1)} M^{(n)^\dagger} F^{(n)^\dagger}
 						\label{rhon}
\eeq
Starting with $\rho^{(0)} = |\nu_j\rangle \langle \nu_j|$ (i.e. $\nu_j$
entering the earth), we can calculate $\rho^{(N)}$ at the
end of the $N$th slab using Eq.(\ref{rhon}). The probability of
observing $\nu_\beta$ at the end of the $N$th slab is
\beq
P^E_{j\beta} = \langle \nu_\beta |\rho^{(N)}|\nu_\beta \rangle  =
(U^{(N)} \rho^{(N)^*} U^{(N)^\dagger})_{\beta\beta} \label{slabprob}
\eeq
This formula (which reduces to Eq.(\ref{pe}) for $N=1$) can be
used for the earth modeled as consisting of $(N+1)/2$
concentric shells, with the density varying gradually
within each shell.

We have already referred to the detection of solar $\nu$ through the
neutral current mode for bypassing the uncertainties of the solar
models. Yet another way would be the detection of the night effect. An
asymmetry between the night and day rates would be an unambiguous
signal for neutrino oscillations independent of the details of the
solar models. The recent results from SK \cite{stanford,learned} for
this asymmetry is at the level of $0.06 \pm 0.03$ and is hence
consistent with zero (at $2\sigma$). Even the absence of the effect
contains important information since it helps to rule out certain
regions of neutrino parameter space in an unambiguous manner.

The night effect is bound to exist at some level and the accummulated
data will soon reveal its magnitude. Further, since the neutrino
samples different amount of matter in the earth during a single night
and also during the period of a year, the data accumulated in various
bins at different times of the night contain an enormous amount of
information on neutrino parameters. We have stressed the importance of
analyzing this time-of-night variation \cite{mohan4} and recent results
from SK \cite{stanford,learned} do suggest such a variation.

Many interesting physical effects are contained in the analytical
formulae already presented. As an example, we shall mention what we may
call "vacuum oscillations in matter".  For $\phi \approx 0$, we get
\cite{mohan4} the following simple formula relating the survival
probability in the night and day :
\beq
P^N_{ee} = P^D_{ee} + \left( 1 - 2 P^D_{ee}\right) \left(P^E_{2e} 
          - \sin^2 \omega\right) \frac{1}{\cos 2\omega}  \label{pnee}
\eeq
where
\bea
P^E_{2e} & = & \sin^2 \omega_E + \sin 2\omega_E \sin 2(\omega_E - \omega)
\sin^2 \frac{1}{4E}
\int^{t_2}_{t_1} \left[m^2_2(t) - m^2_1(t)\right]dt \nn \\
  & \approx & \sin^2 \omega + 2(\omega_E - \omega) \sin 2\omega 
             \sin^2 \frac{\delta m^2_{21} L}{4E}  \label{pe2e}
\eea
Here $\omega_E$ is the mixing angle just below the surface of the earth
and $L$ is the distance the neutrino travels inside the earth. In
arriving at the approximate expression for $P^E_{2e}$ given in
Eq.(\ref{pe2e}), we have assumed that $\delta m^2_{21} >> A (\equiv 2
\sqrt{2} G_F N E)$, $N$ being the electron number density inside the
earth. Under this approximation of small matter effect, $P^E_{2e}$ and
hence $P^N_{ee}$ will exhibit vaccuum type oscillations as a function
of the distance travelled within earth, but their amplitude will be
controlled by matter density (since $(\omega_E - \omega)$ is of order
$A/\delta m^2_{21}$). Such regular oscillations were indicated in the
earlier numerical calculations \cite{raghavan2} for appropriate choice
of parameters and their interpretation is clear from our analytical
formulae. This effect can perhaps be detected at the Borexino (however,
see the remark made above concerning the average over $t_R$).

It is particularly important to see the effect of the core of the earth
\cite{core}. A detector situated near the equator, such as one in South
India \cite{INO} can do this.

\subsection{The eclipse effect}

The above calculation can be extended to include the effect
of the moon  \cite{mohan5}. We can consider both the case of the single
eclipse when the neutrino passes through the moon only and
the case of the ``double eclipse'' when it passes through
the moon and the earth and gets detected on the night-side.
We present the results only. We get \cite{mohan5}, for the single
eclipse,
\beq
P^M_{\alpha \beta} = \sum_j P^S_{\alpha j} P^M_{j\beta}  \label{eclipse1}
\eeq
where
\bea
P^M_{j\beta} &=& \sum_{\ell k \ell ' k'} U^*_{\beta \ell}
U_{\beta \ell '} M^M_{kj} M^{M^*}_{k' j} M^{M^*}_{k\ell} M^M_{k'\ell '}\nn \\  
  & & exp \left\{ -i \left( \epsilon^M_k - \epsilon^M_{k'}\right)
d_M -i\left( \epsilon_\ell - \epsilon_{\ell '}\right)
r\right\}    				\label{pm}
\eea
and for the double eclipse
\beq
P^{ME}_{\alpha \beta} = \sum_j P^S_{\alpha j} P^{ME}_{j\beta}  \label{eclipse2}
\eeq
where
\bea
P^{ME}_{j\beta} &=& \sum_{\stackrel{\ell kp}{\ell ' k' p'}} U^{E^*}_{\beta
p} U^E_{\beta p'} M^E_{p\ell} M^{E^*}_{p'\ell '} M^M_{kj} M^{M^*}_{k' j}
M^{M^*}_{k\ell} M^M_{k' \ell '} \nn \\ 
  & & exp \left\{ -i (\epsilon^M_k - \epsilon^M_{k'}) d_M-i(\epsilon_\ell -
\epsilon_{\ell '}) r-i (\epsilon^E_p - \epsilon^E_{p'}) d_E\right\}
						\label{pme}
\eea
Here, the superscripts $M$ and $E$ refer to the moon and earth
respectively and we have assumed constant densities for simplicity (but
the expressions can be easily generalized to include variable densities
and discrete jumps in densities) ; $d_M$, $d_E$ and $r$ denote the
diameter of the moon, diameter of the earth and the earth-moon distance
respectively ; $M^M$ and $M^{M^*}$ are the non-adiabatic jump
probability amplitudes at the vacuum-moon interface and the moon-vacuum
interface respectively defined analogously to Eq.(\ref{nonad}). 
We again note the
convenient factorization in the results of Eqs(\ref{eclipse1}) and 
(\ref{eclipse2}).

Our calculations \cite{mohan5} show considerable enhancements in the
neutrino counting rate during the eclipse - even as high as 100\%. However,
since the counting rates are currently no more than about one per hour, the
enhancement during the hour or two of the duration of the eclipse is hard to
see at the present detectors. Perhaps we have to wait for the next generation
of detectors.  
\section{Neutrinos from supernovae}
The observation of neutrinos from the supernova SN 1987A was an exciting
event and it spurred much activity in this field. Since we now have
some idea of the mass-differences and the mixing angles from the study
of solar, atmospheric and reactor neutrinos, we may ask what effect do
the oscillations have on the neutrinos from supernovae and whether such
effects can be observed in the neutrino detectors during a supernova
event in the future. Here we shall restrict ourselves to focussing
attention on one such important signal discussed recently \cite{dutta}.

Consider the thermal or cooling phase of the supernova when all the
three flavours of neutrinos and antineutrinos are emitted. We denote the
flux of $\nu_\alpha$ and $\bar{\nu}_\alpha$ produced in the core by
$F^o_\alpha$ and $F^o_{\bar{\alpha}}$ respectively. For all practical
purposes one can put \cite{kuo2}
\beq
F^o_\mu = F^o_{\bar{\mu}} = F^o_\tau = F^o_{\bar{\tau}} \equiv F^o_x
					\label{sourceflux}
\eeq
If $P_{\alpha \beta}$ is the probability for $\alpha$ changing to
flavour $\beta$ during propagation through the supernova, then, the
fluxes of $\nu_e$ and $\nu_\mu + \nu_\tau$ coming out of the supernova
are given by
\bea
F_e &= &F^o_e P_{ee} + F^o_\mu P_{\mu e} + F^o_\tau P_{\tau e} \nn\\
    &=& F^o_e - (1-P_{ee}) (F^o_e - F^o_x)   \nn   \\ 
2F_x &=& F_\mu + F_\tau = 2 F^o_x + (1-P_{ee}) (F^o_e - F^o_x)
						\label{detflux}
\eea
where we have used Eq.(\ref{sourceflux}) and the constraint 
$\sum_\alpha P_{\alpha
\beta} = 1$.
A similar analysis for the antineutrinos gives
\bea
F_{\bar{e}} &=& F^o_{\bar{e}} - (1-P_{{\bar e}{\bar e}}) 
             (F^o_{\bar e} - F^o_{x} )  \nn  \\
2F_{\bar x} &=& 2F^o_x + (1 - P_{{\bar e}{\bar e}}) 
              (F^o_{\bar e} - F^o_x)    \label{nubarflux}
\eea
Let us now calculate $P_{ee}$ and $P_{{\bar e}{\bar e}}$. The variation
of the three matter-dependent mass eigenvalues for the neutrinos and
antineutrinos as functions of matter-density $\rho$ are schematically
depicted in Fig(2). The $\nu_e$ has the decomposition in matter :
\beq
|\nu_e \rangle  = \cos \phi_m \cos \omega_m | \nu^m_1 \rangle  + \cos
\phi_m \sin \omega_m
|\nu^m_2 \rangle  + \sin \phi_m | \nu^m_3 \rangle  \label{nue}
\eeq
where `m' denotes matter. In the dense core of the supernova, $\phi_m
\rightarrow \pi/2$ and so the $\nu_e$ is emitted as $\nu^m_3$ in the
fireball : $ |\nu_e \rangle  = |\nu^m_3 \rangle $.           
\begin{figure}[htbp]
\epsfxsize=12cm
\centerline{\epsfbox{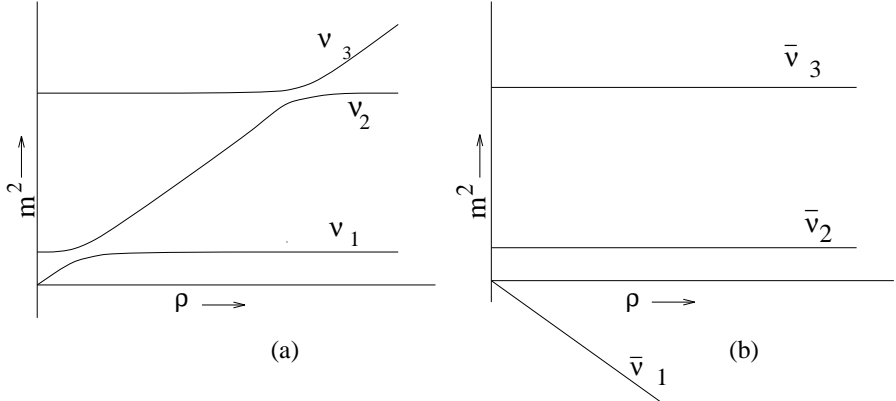}}
\caption{ (a): Mass squares of the three neutrinos as  functions of matter
density $\rho$, with two MSW resonances, (b): Same for antineutrinos where
there are no resonances.} 
\label{fig:lcross3}
\end{figure}
For the parameters relevant for supernovae, one can show \cite{dutta}
that the Landau-Zener nonadiabatic jump probabilities at the two MSW
resonances depicted in Fig(2) are vanishingly small as long as $\sin
\phi \ge 10^{-2}$. Hence, the neutrino state vector $|\nu^m_3\rangle$
evolves adiabatically and ends up as $|\nu_3\rangle $ as it emerges out
of the supernova. Since $\langle \nu_e|\nu_3\rangle  = \sin \phi$, we
have $ P_{ee} = \sin^2 \phi \approx 0 $ where we
have used the result $\phi < 9^o$ from the CHOOZ reactor experiment.

For the antineutrinos, we start with
\beq
|\bar{\nu}_e\rangle  = \cos \phi_m \cos \bar{\omega}_m | \bar{\nu}^m_1 \rangle  + \cos
\phi_m \sin \bar{\omega}_m | \bar{\nu}^m_2\rangle  + \sin \bar{\phi}_m
|\bar{\nu}^m_3\rangle  				\label{nuebar}
\eeq
and since $\bar{\phi}_m \rightarrow 0, \bar{\omega}_m \rightarrow 0$,
at high densities, we see that, when produced, $ |\bar{\nu}_e \rangle
= |\bar{\nu}_1^m \rangle $ and $\bar{\nu}^m_1$ emerges from the
supernova as $\bar{\nu}_1$. Using $\langle \bar{\nu}_e |
\bar{\nu}_1\rangle  = \cos \phi \cos \omega$, we therefore get
$P_{{\bar e}{\bar e}} = \cos^2 \phi \cos^2 \omega \approx 1$ or
${1\over 2}$ for $\phi < 9^o$ and the small or large $\omega$ solar
solution respectively.

Substituting these results into Eqs.(\ref{detflux}) and (\ref{nubarflux}),
we get the changed neutrino fluxes due to oscillations:
\bea
 F_e & \approx & F_x^0  ;~~~2F_x\approx F_e^0 + F_x^0  \label{oflux1} \\
 F_{\bar{e}} & \approx & F_{\bar{x}}^0 ;~~~2F_{\bar{x}}\approx 
                F_{\bar{e}}^0 + F_{\bar{x}}^0  
                   ~~~~~~~~~~~~ \mbox{ (for small}~~~ \omega )
                                                    \label{oflux2}\\
 F_{\bar{e}} & \approx & {1\over 2} (F_{\bar{e}}^0 +F_{\bar{x}}^0) ;~~~
2F_{\bar{x}}\approx {1\over 2}(F_{\bar{e}}^0 + 3F_{\bar{x}}^0)  
                  ~~~~~ \mbox{ (for large} ~~~\omega )   \label{oflux3}
\eea
The changes in the neutrino detection rates arising from these have
been calculated \cite{dutta}, but the important signal for oscillation
is contained in Eq.(\ref{oflux1}) which states that $\nu_x$ ({\it i.e}
$\nu_\mu$ or $\nu_\tau$) are converted into $\nu_e$. Since the original
average energies of $\nu_e$ and $\nu_x$ in the supernova are 12 MeV and
24 MeV, respectively, the average energy of $\nu_e$ is shifted upwards
by the oscillation.  This can be detected by the charged current mode
of $^{16}$O in the H$_2$O detector which has a threshold of 15.4 MeV.
The rate for this mode can be enhanced by as much as two orders of
magnitude as a consequence of oscillation \cite{dutta}. If one can
construct a detector with $^{16}$O or $^{12}$C without protons, it will
be ideal since otherwise the $\bar{\nu}_ep$ absorption reaction is
dominant.  \section{Majorana neutrinos and global analysis} If
neutrinos are Majorana fermions, then neutrinoless double beta decay is
allowed. However the latter has not been seen yet and the experimental
limits on it are getting stronger. The strongest upper limit so far
comes from the Germanium experiment and it is \cite{baudis}
\beq
|\sum_j m_j U^2_{ej} \eta_j | < 0.2 eV  \hskip .5cm {\mbox (at \ 90 \% CL)}
						\label{majconst}
\eeq
where $\eta_j (= \pm 1)$ is the CP parity (apart from a factor $i$) of the
Majorana neutrino $\nu_j$. The mixing matrix for Majorana neutrinos is
\beq
U = \pmatrix{c_\omega c_\phi & s_\omega c_\phi e^{-i \delta_1} & s_\phi 
e^{-i\delta_2} \cr 
-s_\omega c_\psi e^{i\delta_1} -c_\omega s_\psi s_\phi
e^{i(\delta_2 + \delta_3)} & c_\omega c_\psi - s_\omega s_\psi s_\phi
e^{i(\delta_3 + \delta_2 - \delta_1)} & s_\psi c_\phi e^{i\delta_3} \cr
s_\omega s_\psi e^{i(\delta_1 - \delta_3)} - c_\omega c_\psi s_\phi
e^{i\delta_2}
&
-c_\omega s_\psi e^{-i\delta_3} - s_\omega c_\psi s_\phi e^{i(\delta_2 -
\delta_1)}
& c_\psi c_\phi}     \label{majmix}
\eeq
There are three CP-violating phases for Majorana neutrinos, in contrast
to the case of Dirac neutrinos where there is only one phase (see
Eq.\ref{uvac}). However, for oscillation phenomena, only one
combination of the three phases occurs and so oscillations cannot
distinguish between Majorana and Dirac neutrinos.

Is it possible to combine the very important constraint on the neutrino
masses and mixings provided by Eq.(\ref{majconst}) with the information
already derived from the solar, atmospheric and reactor neutrinos ? The
answer is yes, provided we make some assumption about the neutrino mass
scale.  Until two years ago, cosmologists had claimed that their
analysis of the data on the anisotropies of the cosmic microwave
background radiation and the large scale structure of the universe
require the presence of some hot component (presumably massive
neutrinos) in the dark matter and their best fit was \cite{primack}
\beq
\sum^3_{j=1} m_j \approx \ {\rm a \ few} \ eV    \label{summj}.
\eeq
Many years ago \cite{graj}, I had formulated the law that allowed only
a one-way traffic between High Energy Physics and Cosmology : High
Energy Physics $\rightarrow$ Cosmology. We violated this law when we
\cite{rathin} used the cosmological result of Eq.(\ref{summj}) in
neutrino physics and punishment came in the form of the observation
\cite{lsnd} of high red shift supernovae and their interpretation in
terms of a nonvanishing cosmological constant.  The hot dark-matter
component is no longer favoured by cosmologists. So, we now have to
regard Eq.(\ref{summj}) merely as a cosmological assumption about the
neutrino mass scale.

Since the oscillations of the solar and atmospheric neutrinos imply mass
differences which are much smaller than the cosmological scale of 
Eq.(\ref{summj}), we can take
all the three neutrinos as almost degenerate in mass:
$ m_i \approx m_\nu  \approx  1 eV \quad {\mbox (for } \ i = 1,2,3)$
and so Eq.(\ref{majconst}) becomes
\beq
|(\eta_1 \cos^2 \omega + \eta_2 \sin^2 \omega e^{-2i\delta_1}) \cos^2
\phi + \eta_3 \sin^2 \phi e^{-2i\delta_2} | < {0.2}
						\label{etaconst}
\eeq
This constraint can be analysed for all possible
choices of $\delta_1, \delta_2$ and $\eta_i$ 
and the allowed regions for the mixing angles $\omega$
and $\phi$ can be mapped out \cite{rathin}. One fact can be
immediately noted. Since $\phi < 9^o$ according to the
reactor experiment, small values of $\omega$ cannot be
consistent with Eq.(\ref{etaconst}). So we have the important
conclusion : If the cosmological assumption of $m_\nu$ 
in the eV scale is correct and if the
small-$\omega$ solution turns out to be the correct solution of the
solar $\nu$ problem, then neutrinos cannot be Majorana fermions.

\section{LSND and the fourth neutrino}
Since all the results of the solar, atmospheric and reactor neutrino
experiments could be consistently explained within the framework of
three neutrinos, it seemed that all that was required was the
resolution of the three-fold ambiguity of the solar neutrino solutions
and more precision neutrino experiments to pin down the fundamental
neutrino parameters. But a spanner was thrown into the works by the
LSND experiments \cite{athana}
 reporting positive results on $\bar{\nu}_\mu \rightarrow \bar{\nu}_e$
and $\nu_\mu \rightarrow \nu_e$.  Since the base-line length of these
experiments is as short as 29m, the implied $\delta m^2$ is in the
rangle $1-10 eV^2$. It is difficult to incorporate this result within
the three-neutrino framework and so most theorists have decided to
ignore the LSND result, citing the fact that it has not yet been
confirmed by an independent experiment. The independent experiment
KARMEN has not confirmed the LSND result, but KARMEN \cite{karmen} has
not ruled out the full parameter space allowed by LSND either. A real
confirmation or ruling out has to await the Mini BOONE experiment at
FNAL, a long agonizing 4 years away.

If the LSND result is correct, we need a 4th neutrino, but since the
known invisible width of $Z$ is completely exhausted by $\nu_e,
\nu_\mu$ and $\nu_\tau$, the new neutrino has to be a singlet under
$SU(2)$ and be sterile under known interactions. The natural mass
hierarchy would be to place $\nu_4$ a few eV above the known three
neutrinos, but this is contradicted \cite{okada} by a combination of
known experimental data, unless $\nu_4$ decays \cite{ma}.

\section{Future}
Is it possible to confirm or refute the results on neutrino
oscillations claimed by the solar and atmospheric neutrino observations
using laboratory experiments ?  That would be one of the chief goals of
the long-base-line neutrino experiments that are being planned (see
Table 1.) One can see that sensitivities upto the level of $\Delta m^2$
needed for solar and atmospheric neutrinos will be reached in these
experiments.  Even larger baselines can be contemplated.
\begin{table}[h]

\begin{tabular}{|l|l|c|c|c|l|}   & & & & & \\
~Expt & Baseline & $L(km)$ & $\langle E_\nu\rangle GeV$ & 
                                          $\Delta m^2 (eV^2)$& Status \\ 
&  &  & & probed &  \\ \hline & & & & & \\
~K2K & KEK $\rightarrow$ Kamioka & 250 & 1.4 & 10$^{-3}$ & taking data \\
& FNAL $\rightarrow$ Soudan & 730 & 10 & 10$^{-2}$ & to start in 2002 \\
& CERN $\rightarrow$ Gransasso & 732 & 20 & 10$^{-2}$ & to start in 2005 \\
~Kamland & Reactors $\rightarrow$ Kamioka & 160 & $3\times 10^{-3}$
            & 10$^{-5}$ & 
                                                    to start in 2001 
  
\end{tabular}

\hskip .5cm

\caption {Long-base-line neutrino experiments}
\end{table}
Finally, there are prospects of constructing muon
storage rings that will function as neutrino factories and these promise 
to take neutrino physics to a new era. Hopefully these as well as the long
baseline experiments will lead to a determination of the neutrino parameters. 
Of course, entirely new phenomena could also be discovered.


\begin{thebibliography}{99}
\bibitem{davis1} R. Davis, D. S. Harmer and K. C. Hoffman, Phys. Rev. Lett. 
  {\bf 20}, 1205 (1968).
\bibitem{bahcall1} J. N. Bahcall, {\it Neutrino Astrophysics}
  (Cambridge Univ. Press, 1989)
\bibitem{pakvasa} S. Pakvasa, Talk at this Workshop.
\bibitem{mohan1} M. Narayan, M. V. N. Murthy, G. Rajasekaran and S.
  Uma Sankar, Phys. Rev. {\bf D53}, 2809 (1996).
\bibitem{mohan2} M. Narayan, G. Rajasekaran and S. Uma Sankar, Phys. Rev. {\bf
  D 56}, 437 (1997).
\bibitem{chooz} CHOOZ Collaboration, M. Apollonio {\it et al}, Phys. Lett. {\bf
  B420}, 397 (1998).
\bibitem{mohan3} M. Narayan, G. Rajasekaran and S. Uma Sankar, Phys. Rev. {\bf
  D58}, 031301 (1998).
\bibitem{parkes} S. Parke, Phys. Rev. Lett., {\bf 57}, 1275 (1986).
\bibitem{kuo1} T. K. Kuo and J. Pantaleone, Rev. Mod. Phys., {\bf 61}, 937
  (1989).
\bibitem{fogli1} G. L. Fogli, E. Lisi and D. Montanino, Phys. Rev. {\bf D49},
  3626 (1994); G. L. Fogli and E. Lisi, Astroparticle Phys. {\bf 3}, 185
  (1995).
\bibitem{fogli2} G. L. Fogli, E. Lisi, D. Montanino and G. Scioscia, 
  Phys. Rev. {\bf D55}, 4385 (1997);
  G. L. Fogli, E. Lisi, A. Marrone and G. Scioscia, hep-ph/9808205;
  S. Goswami, K. Kar and A. Raychaudhury, Int. J. mod. phys. {\bf A12}, 781
  (1997); S. Goswami, Phys. Rev. {\bf D55}, 2931 (1997); 
  S. M. Bilenky, C. Giunti and C. W. Kim, hep-ph/9505301. 
\bibitem{davis2} R. Davis, Prog. Part. Nucl. Phys. {\bf 32}, 13 (1994); 
  P. Anselman {\it et al.}, Phys. Lett. {\bf B357}, 237 (1995); {\bf B361}, 235
  (1996); J. N. Abdurashitov {\it et al}, Phys. Lett. {\bf B328}, 234 (1994);
  Y. Fukuda {\it et al.}, Phys. Rev. Lett. {\bf 77}, 1683 (1996); {\bf 81}, 1158
  (1998); {\bf 82}, 1810 (1999); {\bf 82}, 2430 (1999).
\bibitem{bahcall2} J. N. Bahcall, P. I. Krastev and A. Yu. Smirnov,
   hep-ph/9807216 
\bibitem{donald} A. B. Mc Donald, in Particle Phys. and Cosmology, Proc. of
  the 9th. Lake Louise Winter Institute, 1994, edited by A. Astbury {\it et al}, 
  (World Scientific Singapore, 1995), p.1.
\bibitem{raghavan1} R. S. Raghavan, Science {\bf 267}, 45 (1995). 
\bibitem{fukuda} Y. Fukuda {\it et al}, Phys. Lett. {\bf B433}, 9 (1998); {\bf
  B436}, 33 (1998); Phys. Rev. Lett. {\bf 81}, 1562 (1998); {\bf 82}, 2644
  (1999).
\bibitem{geophysical} R. S. Raghavan {\it et al.}, Phys. Rev. Lett. {\bf 80},
  635 (1998).
\bibitem{mohan4} M. Narayan, G. Rajasekaran and R. Sinha, Mod. Phys. Lett. 
  {\bf A13}, 1915 (1998).
\bibitem{core} J.M. Gelb, W. Kwong, and S.P. Rosen, Phys. Rev. Lett.
{\bf 78}, 2296 (1997).
\bibitem{INO} Discussion on a possible Indian Neutrino Observatory
  at this Workshop. 
\bibitem{mohan5} M. Narayan, G. Rajasekaran, R. Sinha and C. P. Burgess,
  Phys. Rev. {\bf D60}, 073006 (1999).
\bibitem{stanford} Y. Suzuki, Talk at XIX International Symposium on 
   Lepton and Photon Interactions at High Energies, Stanford, 1999. 
\bibitem{learned} J. Learned, Talk at this Workshop.  
\bibitem{raghavan2} R. S. Raghavan, A. B. Balantekin, F. Loreti, A. J. Baltz,
  S. Pakvasa and J. Pantaleone, Phys. Rev. {\bf D44}, 3786 (1991). 
\bibitem{dutta} G. Dutta, D. Indumathi, M. V. N. Murthy and G. Rajasekaran,
  Phys. Rev. {\bf D61}, 013009 (1999).	 
\bibitem{kuo2} T. K. Kuo and J. Pantaleone, Phys. Rev. {\bf D37}, 298 (1988). 
\bibitem{baudis} L. Baudis et al., Phys. Rev. Lett. {\bf 83}, 41 (1999).
\bibitem{primack} J. R. Primack, Phys. Rev. Lett. {\bf 74}, 2160 (1995);
  E. Gawiser and J. Silk, Science {\bf 280}, 1405 (1998); 
  J. R. Primack, {\it ibid}. {\bf 280}, 1398 (1998).
\bibitem{graj} G. Rajasekaran, in Proc. of VIII HEP Symposium, Saha Institute 
  of Nucl. Phys., 1986(Vol II) 
  Ed. M. K. Pal and G. Bhattacharya, p 399. 
\bibitem{rathin} R. Adhikari and G. Rajasekaran, Phys. Rev. {\bf D61}, 
  031301 (1999).
\bibitem{lsnd} S. Perlmutter {\it et al}, Nature (London) {\bf 391}, 51 (1998).
\bibitem{athana} C. Athanassopoulos et al., Phys. Rev. Lett. {\bf 75}, 2650
  (1995);{\bf 77}, 3082 (1996); {\bf 81}, 1774 (1998).
\bibitem{karmen}T. E. Jannakos, hep-ex/9908043; B. Zeitnitz {\it et al.}, 
  Prog. Part. Nucl. Phys. {\bf 32}, 351 (1994). 
\bibitem{okada} N. Okada and O. Yasuda, Int. J. Mod. Phys. {\bf A12}, 3669
  (1997); S. M. Bilenky, C. Giunti and W. Grimus, Eur. Phys. J. {\bf C1},
  247 (1998); V. Barger, S. Pakvasa, T. J. Weiler and K. Whisnant, Phys. Rev.
  {\bf D58}, 093016 (1998).
\bibitem{ma}E. Ma, G. Rajasekaran and I. Stancu, Phys. Rev. {\bf D61},
  071302(R) (2000).
\end{thebibliography}
\end{document}